\documentclass[10pt]{article}
\usepackage{fa2025}
\usepackage{amsmath}
\usepackage{cite}
\usepackage{url}
\usepackage{graphicx}
\usepackage{color}
\usepackage{siunitx}
\usepackage{acronym}
\usepackage{amsfonts}

\usepackage[T1]{fontenc}

\acrodef{RIR}{Room Impulse Response}
\acrodef{CD}{Cosine Distance}
\acrodef{NMSE}{Normalized Mean Square Error}

\acrodef{GAN}{Generative Adversarial Network}
\acrodef{CNN}{Convolutional Neural Network}
\acrodef{DDPM}{Denoising Diffusion Probabilistic Models}
\acrodef{PINN}{Physics-informed Neural Network}
\acrodef{EDC}{Energy Decay Curve}
\acrodef{ULA}{uniform linear array}
\acrodef{SCI}{Spline Cubic Interpolation}

\title{DiffusionRIR: Room Impulse Response Interpolation using Diffusion Models}

\multauthor
{Sagi Della Torre$^{1}$ \hspace{1cm} Mirco Pezzoli$^2$ \hspace{1cm} Fabio Antonacci$^2$ \hspace{1cm} Sharon Gannot$^{1*}$} { \\
$^1$ Faculty of Engineering, Bar-Ilan University, Israel\\
$^2$ Dipartimento di Elettronica, Informatica e Bioingegneria, Politecnico di Milano, Italy
\correspondingauthor{sharon.gannot@biu.ac.il}{Sagi Della Torre et al.}
}

\begin{document}

\maketitle
\begin{abstract}
\acp{RIR} characterize acoustic environments and are crucial in multiple audio signal processing tasks. High-quality \ac{RIR} estimates drive applications such as virtual microphones, sound source localization, augmented reality, and data augmentation. However, obtaining \ac{RIR} measurements with high spatial resolution is resource-intensive, making it impractical for large spaces or when dense sampling is required. This research addresses the challenge of estimating \acp{RIR} at unmeasured locations within a room using \ac{DDPM}. Our method leverages the analogy between \ac{RIR} matrices and image inpainting, transforming \ac{RIR} data into a format suitable for diffusion-based reconstruction.

Using simulated \ac{RIR} data based on the image method, we demonstrate our approach's effectiveness on microphone arrays of different curvatures, from linear to semi-circular. Our method successfully reconstructs missing \acp{RIR}, even in large gaps between microphones. Under these conditions, it achieves accurate reconstruction, significantly outperforming baseline \ac{SCI} in terms of \ac{NMSE} and \ac{CD} between actual and interpolated \acp{RIR}.

This research highlights the potential of using generative models for effective \ac{RIR} interpolation, paving the way for generating additional data from limited real-world measurements.
\end{abstract}

\section{Introduction}\label{sec:introduction}

\acfp{RIR} play a critical role in audio signal processing, enabling applications such as sound source localization, virtual and augmented reality, and data augmentation for machine learning. However, measuring \acp{RIR} is resource-intensive, particularly in large or acoustically complex spaces requiring dense measurements. Simulated \acp{RIR}, while practical, often lack the accuracy and fidelity of real-world data, necessitating methods to reconstruct or interpolate \acp{RIR} at unmeasured locations.

Traditional methods for \ac{RIR} reconstruction rely on mathematical models, such as compressed sensing and wave equation solutions \cite{thiergart2013geometry, pezzoli2020parametric, pezzoli2022sparsity, fahim2017sound, zea2019compressed}, but these approaches often struggle with complex acoustic environments. Recent advancements leverage deep learning techniques, including \acp{CNN} \cite{pezzoli2022deep} and \acp{GAN}, to improve reconstruction accuracy. For instance, \acp{GAN} have shown promise in extending the bandwidth of array processing \cite{fernandez2023generative}, while \acp{PINN} incorporate acoustic principles to refine predictions \cite{olivieri2021physics}. \ac{DDPM} has recently emerged as a powerful tool for sound field reconstruction, offering a probabilistic framework for generating accurate acoustic fields \cite{miotello2024reconstruction}. However, most of these approaches focus on specific frequency bands or parts of the \ac{RIR}. A recent challenge focuses on generative models for synthesizing room acoustics as a data augmentation tool for speaker distance estimation tasks \cite{lin2025generative}. 

Our work explores the analogy between \ac{RIR} reconstruction and image inpainting. By treating \ac{RIR} matrices as images, we apply a diffusion model to reconstruct the full time span of \acp{RIR}. This novel approach enables robust and accurate \ac{RIR} interpolation, achieving excellent performance in terms of \ac{NMSE} and \ac{CD}, even in scenarios where the microphones are sparsely distributed in the acoustic environment. The proposed method, supported by an experimental study using simulated acoustic environments, provides a strong foundation for potential real-world applications.

\section{Problem Formulation}

This research aims to reconstruct \acp{RIR} for unmeasured locations using a limited number of measured \acp{RIR}. Given $M$ measured \acp{RIR} in a room, the task is to estimate \acp{RIR} at $L$ unmeasured locations, resulting in a total of $N = M + L$ locations. Each \ac{RIR} is sampled at a frequency $F_s$ and truncated to $K$ samples, beyond which it falls into the noise floor.

\begin{figure}[ht]
 \centerline{\framebox{
 \includegraphics[width=7.8cm]{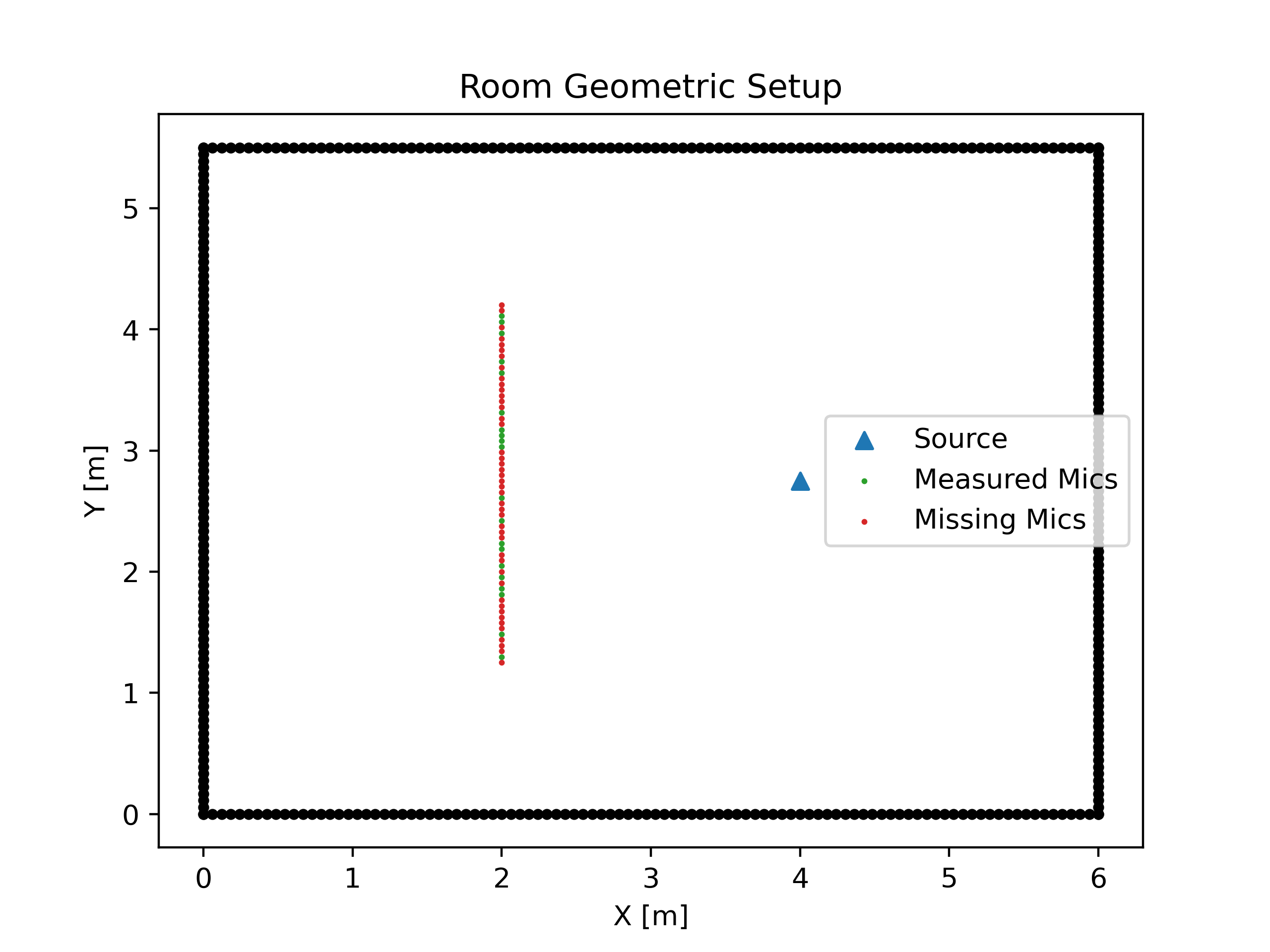}}}
 \caption{Geometric setup of a room with a source and a microphone array. Measured and missing microphones are marked in green and red dots, respectively. We aim to reconstruct the \acp{RIR} of the missing microphones.}
 \label{fig:geometric_setup}
\end{figure}

This paper focuses on linear and semi-circular array configurations, as well as intermediate arc-shaped configurations, although the methodology can be extended to other setups. In this framework, we consider $N$ microphone positions, of which only $M$ randomly selected \acp{RIR} are measured, while the remaining $L$ measurements are missing, as illustrated in Fig.~\ref{fig:geometric_setup} for a linear array.

\begin{figure}[ht]
 \centerline{\framebox{
 \includegraphics[width=7.8cm]{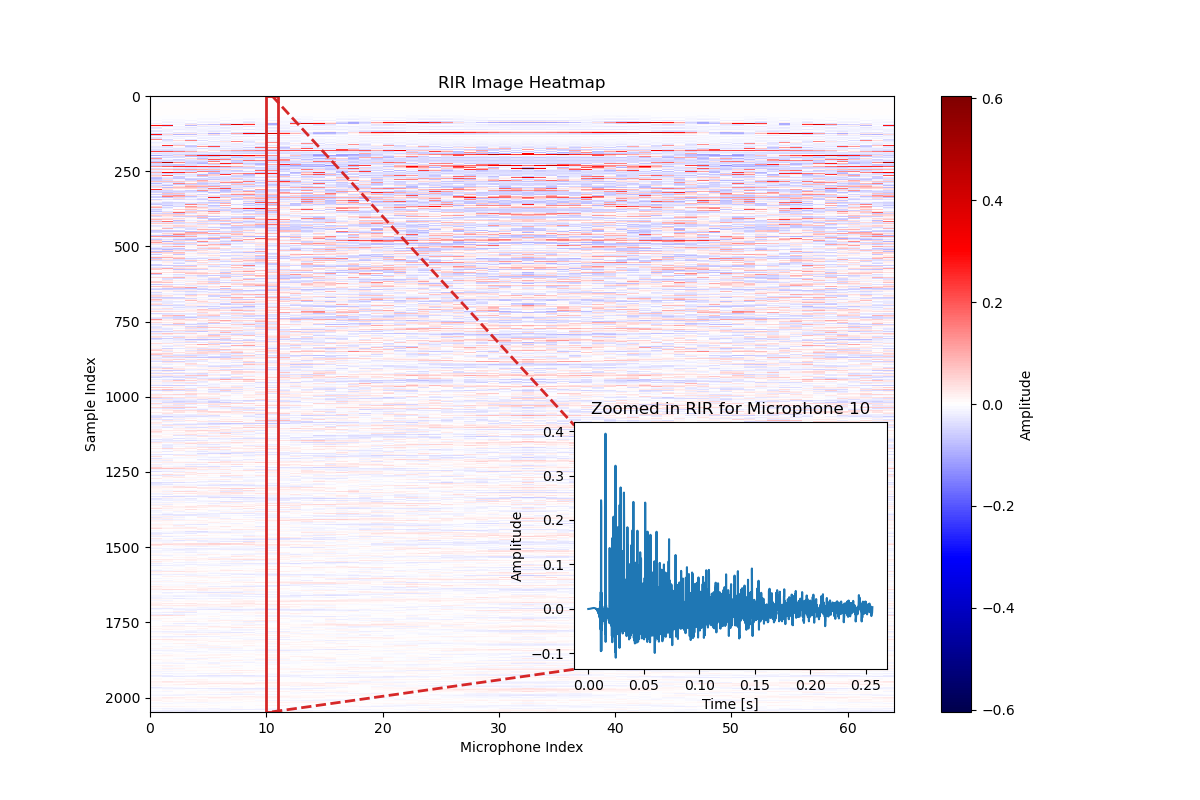}}}
 \caption{Heatmap of RIR matrix \(\mathbf{H}\), with one zoomed-in RIR view. Each column in the matrix represents one microphone.}
 \label{fig:heatmap}
\end{figure}

Mathematically, let $\mathbf{H}$ be the matrix representing the \acp{RIR}, where $\mathbf{H} \in \mathbb{R}^{N \times K}$. We denote the available \ac{RIR} measurements as $\mathbf{H}_{\text{measured}} \in \mathbb{R}^{M \times K}$. Our objective is to estimate the missing entries in $\mathbf{H}$ to obtain a complete matrix $\hat{\mathbf{H}} \in \mathbb{R}^{N \times K}$. Each column of $\mathbf{H}$, denoted $\mathbf{h}_i$, represents the \ac{RIR} at the $i$-th location, where $1 \leq i \leq N$. Treating this matrix as an image, the problem is analogous to image inpainting, where the goal is to reconstruct the missing parts using the available data. Figure \ref{fig:heatmap} shows a heatmap of the matrix $\mathbf{H}$, along with a zoomed-in view of one microphone's \ac{RIR}. Our aim is to reconstruct the missing \acp{RIR} scattered throughout the array using the measured \acp{RIR}.
Reconstructing missing \acp{RIR} requires leveraging the data's spatial and temporal structures. By addressing this challenge, we aim to develop a robust interpolation method that facilitates acoustic analysis and processing across various applications.

\section{Proposed Approach}

We formulate the problem of reconstructing missing \acp{RIR} as an image inpainting task. By representing the \ac{RIR} data as an image, we can leverage the power of \acp{DDPM} to estimate the missing responses. Our work is inspired by previous research on image inpainting using diffusion models, notably \cite{lugmayr2022repaint}, which demonstrated effective reconstruction of missing image regions using a pretrained diffusion model which was trained on the task of generating new images.

\subsection{Inpainting with Diffusion Models}

Lugmayr et al.~\cite{lugmayr2022repaint} introduced RePaint, an inpainting method based on \acp{DDPM}. This approach utilizes a pre-trained model originally trained for general image generation. During inference, the model is adapted to the inpainting task by conditioning it on the known parts of an image while generating new content for the missing regions. At each diffusion step, the model is guided to remain consistent with the observed parts, ensuring that only the missing regions are reconstructed while the known areas are preserved. This method allows for flexible inpainting without requiring prior knowledge of the mask pattern.

This iterative refinement aligns well with our problem, where missing \ac{RIR} data should be reconstructed to closely resemble the original responses without prior knowledge of the missing microphone positions.

We adopt OpenAI's \ac{DDPM} architecture\footnote{\url{https://github.com/openai/guided-diffusion}} with necessary modifications to accommodate \ac{RIR}-matrix images. While the original model is designed for natural images, \ac{RIR} data exhibits distinct statistical properties. Training the model on a dedicated small \ac{RIR} dataset allows it to capture these characteristics, leading to more accurate reconstructions.

During inference, a masked \ac{RIR} image is fed into the trained diffusion model, which iteratively reconstructs the missing regions. The output is a complete \ac{RIR} image. Finally, the reconstructed image is converted to its original matrix form by transforming grayscale pixel values into response amplitudes. Only the newly inpainted regions are retained, representing the reconstructed \ac{RIR}.

\subsection{Image Representation of the \ac{RIR} Set}

To apply inpainting techniques, we recast the \ac{RIR} data into an image-like format. Given an array configuration, we arrange the \acp{RIR} into a 2D matrix where each column corresponds to an \ac{RIR} of length $K$ from a specific microphone position. Different numbers of missing microphones and \ac{RIR} lengths can also be accommodated. This will result in images of varying width and height dimensions. The resulting matrix is treated as a grayscale image, with intensity values representing normalized \ac{RIR} amplitudes. This format enables structured processing while retaining spatial and temporal information.

Since \acp{DDPM} are typically trained on fixed-size images, we split the \ac{RIR} matrix into patches of $64 \times 64$ pixels, corresponding to 64 possible microphone positions and 64 \ac{RIR} taps. If the length of the \ac{RIR} exceeds 64, as is often the case, we divide the image into multiple patches, each representing a different portion of the \ac{RIR}. 

To address the issue of lower reconstruction quality at the edges of the patches due to the lack of surrounding context, we introduce an overlap of 25\% between adjacent patches. We also normalize each patch to the range -1 to 1, allowing the network to reconstruct each patch independently of the energy level of that part of the response. After reconstruction, these patches are reassembled into a complete image by rescaling each patch to its original energy, discarding the overlapping regions, and retaining only the central portions of the patches. This approach balances computational efficiency and reconstruction accuracy and ensures a seamless reconstruction by eliminating duplicates and maintaining continuity. 

In cases where the microphone configuration has fewer than 64 microphones, we pad the image with duplicated columns to ensure an image width of 64 pixels. This preserves the model's expected input dimensions while minimizing distortions in the reconstruction process.  

To simulate missing measurements, we generate masks of varying percentages by zeroing out randomly selected columns in the \ac{RIR} image. These masks represent the unmeasured microphone locations. The masked image, along with its corresponding mask, are then fed as input to the diffusion model.

\section{Experiments}
In this section, we describe our experiments using artificial \acp{RIR} generated by the Pyroomacoustics package.\footnote{\url{https://pyroomacoustics.readthedocs.io/en/pypi-release/}}

\subsection{Experiment Setup}

The simulated database of \acp{RIR} comprises multiple microphone array configurations: a \ac{ULA}, a semi-circular array, and intermediate arcs. The \ac{ULA} configuration uses 64 microphones and spans a length of 3 meters, resulting in a 4 cm distance between adjacent microphones. The semi-circular array configuration uses 64 microphones with a 1.5-meter radius, resulting in a 7.3 cm distance between adjacent microphones. The simulated room dimensions are \(6 \times 5.5 \times 2.8 \, \text{m}\) (length, width, and height, respectively). The simulation uses a sampling frequency of $F_s=8$~kHz.

The data is split into training and inference sets.
The training set consists of 176 randomly selected patches from 8 \ac{RIR} images, corresponding to various microphone array configurations. The source positions were randomly selected from 9 positions on a semi-circle with a radius of 2 meters from the array's center, as depicted in Fig.~\ref{fig:room_setup_different_source}. During training, the reverberation time (${T_{60}}$) was fixed at 0.3 seconds across all frequency bands. Each \ac{RIR} was truncated to 1024 samples, corresponding to a duration of 0.128 seconds.

As previously mentioned, during training, the model learns to generate new images from the distribution of the training dataset. During inference, the model generates new images while conditioning on the known parts of the image, which correspond to the measured responses.

To introduce variability in absorption coefficients across frequency bands, we selected ``smooth brickwork 10~mm pointing,'' from the Pyroomacoustics material database as the wall material for the inference dataset. Using this material and the specified room dimensions corresponds to a full-band reverberation time (${T_{60}}$) of 0.6 seconds. The evaluation was carried out using a \ac{ULA}, a semi-circular array, and intermediate arc configurations. All nine different source positions were tested. Each \ac{RIR} was truncated to 2048 samples, corresponding to a duration of 0.256 seconds.
In each trial, we randomly removed a percentage of the microphone measurements, varying the ratio of missing microphones from 10\% to 90\%.

\subsection{Performance Measures and a Baseline Method}
The results of our experiments are analyzed using two quality measures, comparing the estimated and the ground truth \ac{RIR} for the $M$ missing microphones.
The first is the \acf{NMSE} in dB, defined as (see \cite{zea2019compressed}):
\begin{equation}
\text{NMSE}(\mathbf{H}, \hat{\mathbf{H}}) = 10 \log_{10} \left( \frac{1}{M} \sum_{i=1}^{M} \frac{\|\hat{\mathbf{h}}_i - \mathbf{h}_i\|^2}{\|\mathbf{h}_i\|^2} \right),
\end{equation}
where $\hat{\mathbf{h}}_i \in \mathbb{R}^{N \times 1}$ is the estimate of the $i$th \ac{RIR} corresponding to the $i$th column of $\hat{\mathbf{H}}$.
The second is the \acf{CD}, defined as (see also \cite{morgan1998evaluation}):
\begin{equation}
\text{{CD}}(\mathbf{H}, \hat{\mathbf{H}}) = \frac{1}{M} \sum_{i=1}^{M} \left( 1 - \left( \frac{\mathbf{h}_i^\top \hat{\mathbf{h}}_i}{\|\mathbf{h}_i\| \|\hat{\mathbf{h}}_i\|} \right)^2 \right).
\end{equation}
The value $\text{{CD}}(\mathbf{H}, \hat{\mathbf{H}}) = 1$ is obtained if all estimates are orthogonal to the corresponding true \ac{RIR} for all $M$ missing values (i.e., all estimates are the worst possible), and $\text{{CD}}(\mathbf{H}, \hat{\mathbf{H}}) = 0$ if all estimated and true \acp{RIR} are perfectly aligned.   
The \ac{CD} is particularly useful in audio applications \cite{morgan1998evaluation}.

Finally, we used the \acf{SCI} technique as a baseline method \cite{de1978practical}.

\subsection{Results}
In this section, we present and analyze the performance of the proposed method and compare it with the baseline method.

\begin{figure}[htb]
 \centerline{\framebox{
 \includegraphics[width=7.8cm]{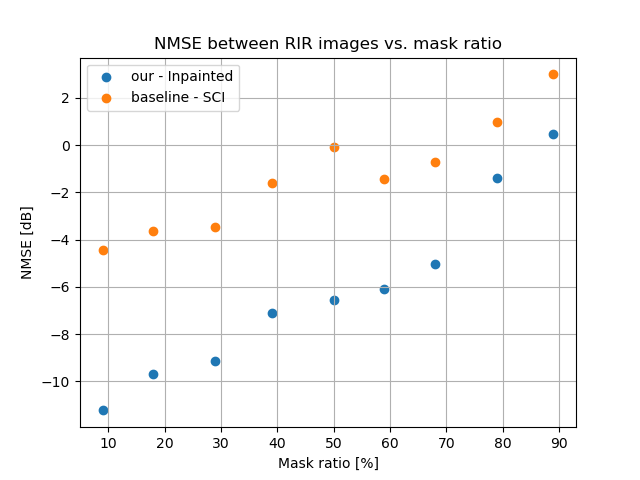}}}
 \caption{\ac{NMSE} for inpainted and baseline \ac{SCI} vs. mask ratio.}
 \label{fig:nmse_error}
\end{figure}

\begin{figure}[htb]
 \centerline{\framebox{
 \includegraphics[width=7.8cm]{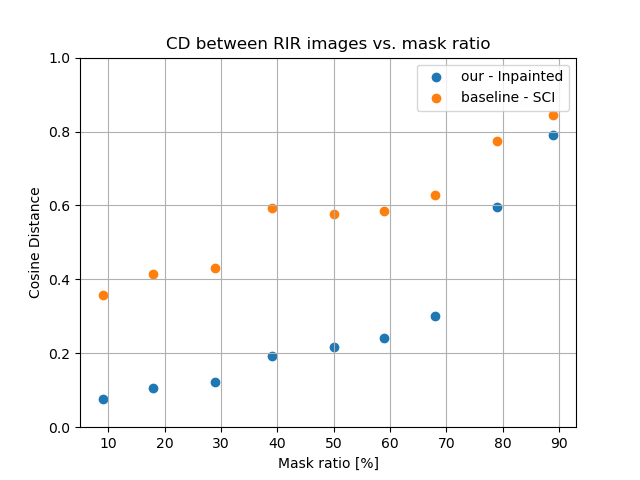}}}
 \caption{\ac{CD} for inpainted and baseline \ac{SCI} vs. mask ratio.}
 \label{fig:npm_error}
\end{figure}

Figures \ref{fig:nmse_error} and \ref{fig:npm_error} depict the \ac{NMSE} and \ac{CD}, respectively, for different mask ratios for linear array configuration as shown in Fig.~\ref{fig:geometric_setup}.
Our method improves the \ac{NMSE} by 3 to 7 dB and the \ac{CD} measure over the baseline by approximately 0.3, depending on the mask ratio.

\begin{figure}[ht]
 \centerline{\framebox{
 \includegraphics[width=7.8cm]{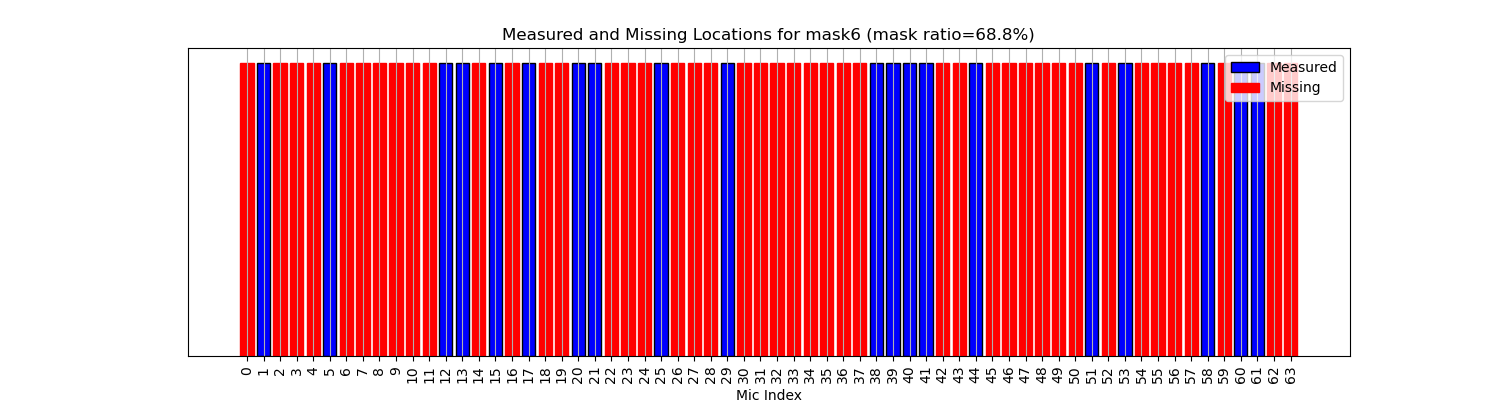}}}
 \caption{Linear array configuration: Measured (blue) and missing (red) microphones.}
 \label{fig:missing_location}
\end{figure}

\begin{figure}[h!]
 \centerline{\framebox{
 \includegraphics[width=7.8cm]{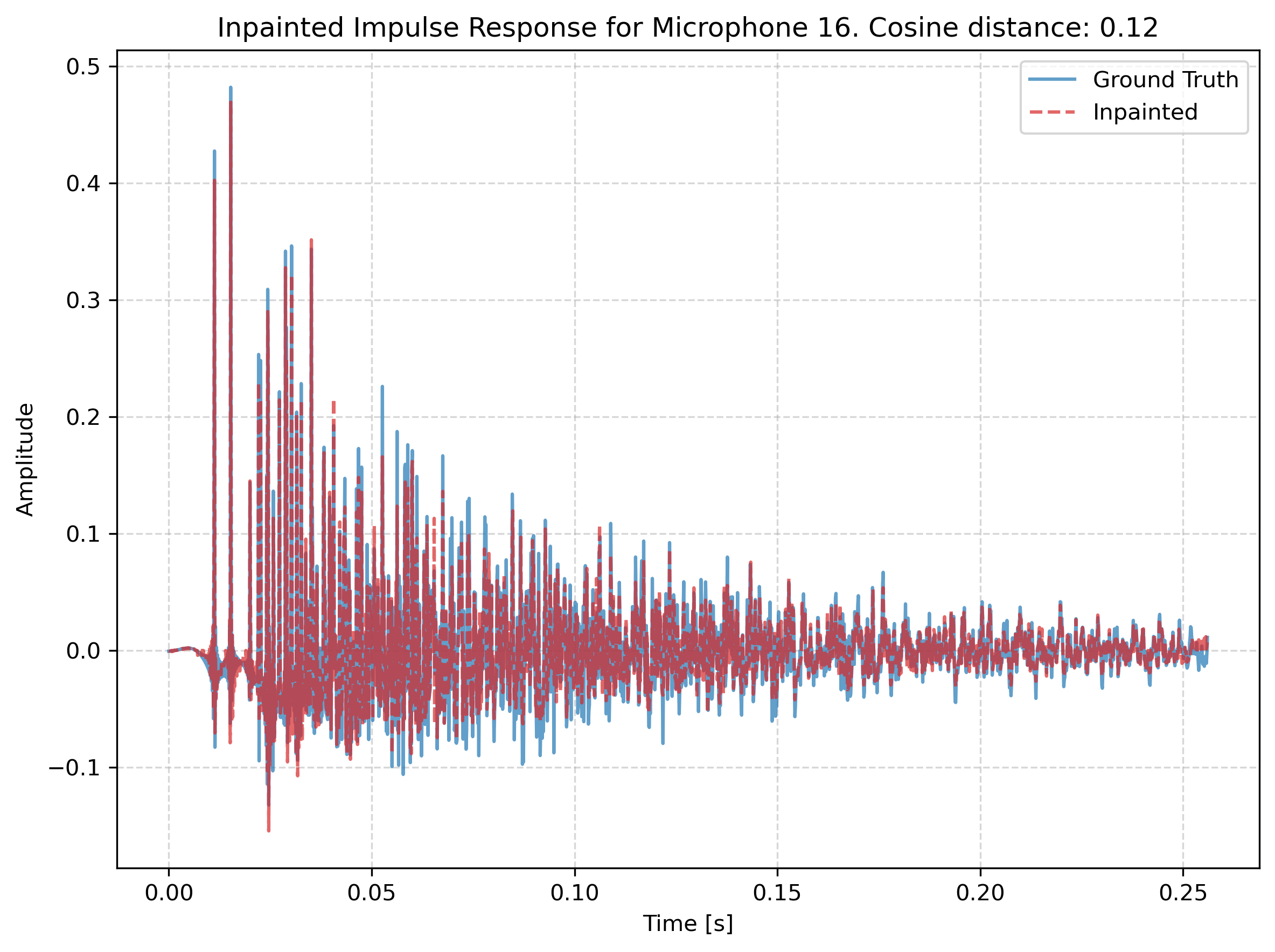}}}
 \caption{Reconstructed and ground truth impulse response for microphone No. \#16.}
 \label{fig:reconstructed_rir_16}
\end{figure}

\begin{figure}[h!]
 \centerline{\framebox{
 \includegraphics[width=7.8cm]{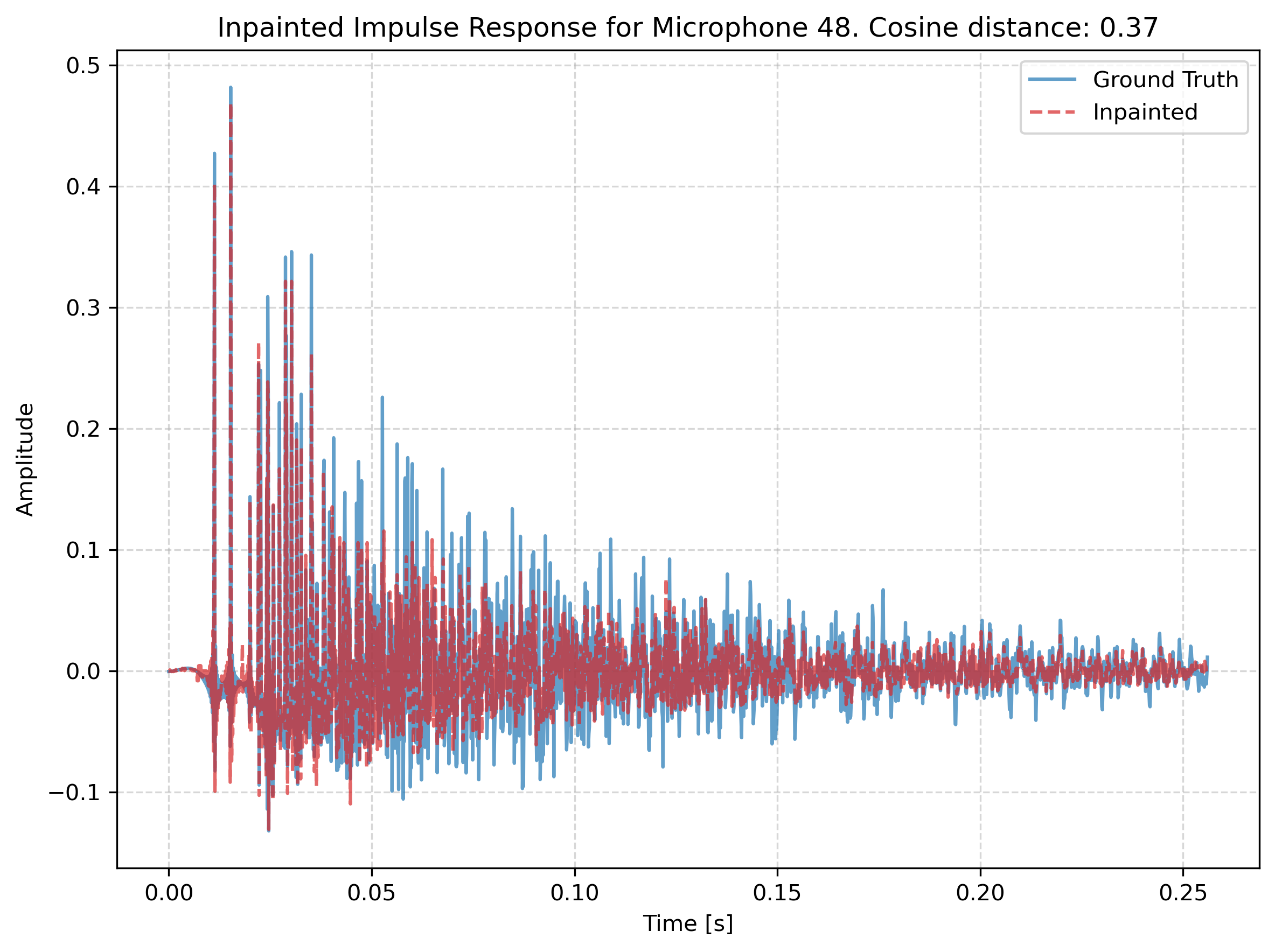}}}
 \caption{Reconstructed and ground truth impulse response for microphone No. \#48.}
 \label{fig:reconstructed_rir_48}
\end{figure}

\begin{figure}[h!]
 \centerline{\framebox{
 \includegraphics[width=7.8cm]{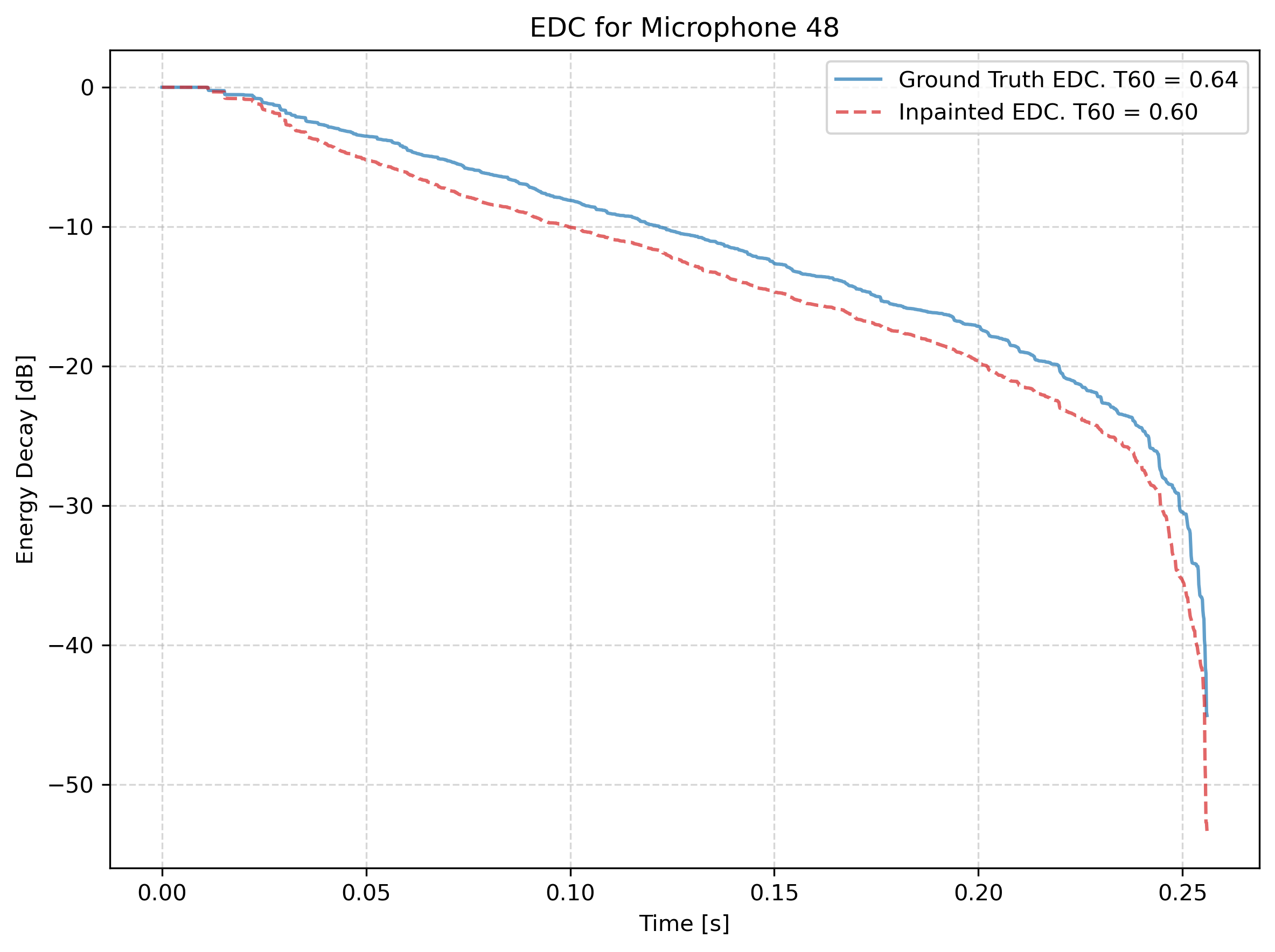}}}
 \caption{\ac{EDC} of reconstructed and ground truth impulse response for microphone No. \#48, and the corresponding $T_{60}$.}
 \label{fig:edc_48}
\end{figure}

We now focus on the reconstruction results for a 70\% mask ratio. Figure~\ref{fig:missing_location} provides a detailed view of the microphone indices, describing which microphone signals are measured (blue) and which are missing (red). 
In Figs.~\ref{fig:reconstructed_rir_16} and \ref{fig:reconstructed_rir_48}, we present the reconstructed and ground truth \acp{RIR} for microphones \#16 and \#48, respectively. Microphone \#16 is located very close to the measured microphones, while microphone \#48 is situated in a region with sparse measurements, leading to better reconstruction for the former. Yet, even in the more challenging case of microphone \#48, the reconstructed \ac{RIR} successfully captures the main features, including both the direct and early arrivals. The \ac{CD} for microphone \#48 is relatively low at 0.37 (but higher than $\text{CD} = 0.12$ for microphone \#16), demonstrating the model's ability to accurately infer and reconstruct acoustic reflections even in areas with limited measured data.
In Fig.~\ref{fig:edc_48}, we further analyze the acoustic properties of \ac{RIR} \#48 by examining its \ac{EDC}. It is evident that the \ac{EDC} of the reconstructed \ac{RIR} closely resembles that of the ground truth. Moreover, the estimated full-band reverberation time derived from the \ac{EDC} slope, ${T_{60}}=0.64$ seconds, closely matches the ground truth value of ${T_{60}}=0.6$ seconds.
Despite the large percentage of missing microphones, our method demonstrates favorable performance for the \ac{ULA} configuration, generating a reconstructed \acp{RIR} that closely align with the ground truth responses.

\begin{figure}[h!]
 \centerline{\framebox{
 \includegraphics[width=7.8cm]{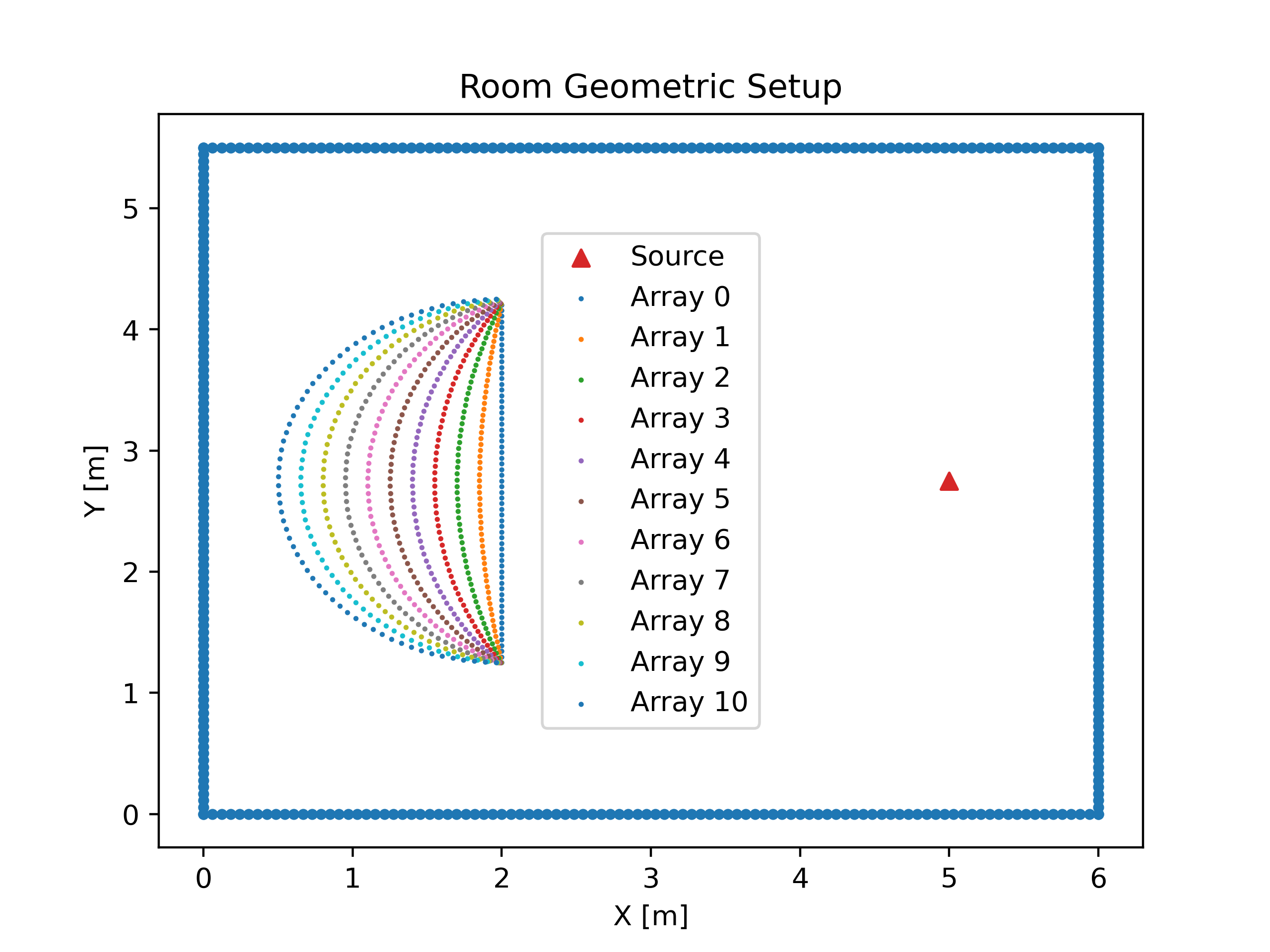}}}
 \caption{Room with several array curvatures from a linear to a semi-circular configuration.}
 \label{fig:linear_to_circular_geometric_setup}
\end{figure}

Next, we evaluate the performance of the proposed method for the semi-circular microphone array. First, Fig.~\ref{fig:linear_to_circular_geometric_setup} illustrates the room and several array curvatures. We begin by examining the semi-circular array labeled as array \#10 in the figure.

\begin{figure}[h!]
 \centerline{\framebox{
 \includegraphics[width=7.8cm]{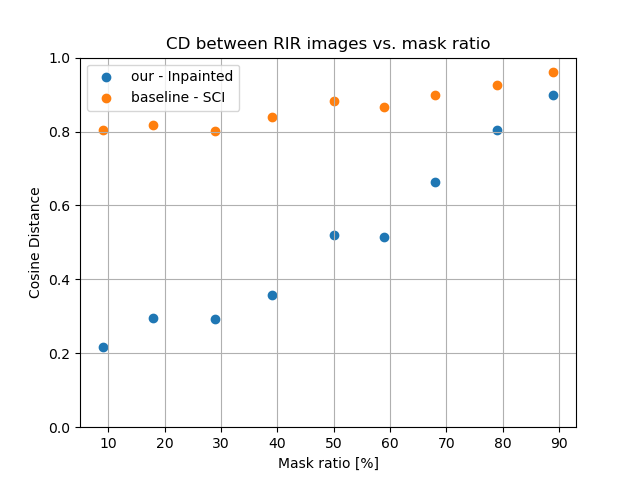}}}
 \caption{\ac{CD} for semi-circular array vs. mask ratio.}
 \label{fig:npm_errors_circular_array}
\end{figure}

The reconstruction results for the semi-circular array, as measured by the \ac{CD} metric, are presented in Fig.~\ref{fig:npm_errors_circular_array}. Our method significantly outperforms the baseline interpolation approach, with improvements of 0.2–0.6 in the \ac{CD} measure, up to a mask ratio of 70\%.

Despite these improvements, the inpainting algorithm's performance for the semi-circular array is inferior compared to the linear array. This difference can be attributed to the geometric challenges posed by the curved reflection patterns in the semi-circular array, which are more complex to reconstruct than the straight-line patterns found in the linear array.

\begin{figure}[h!]
 \centerline{\framebox{
 \includegraphics[width=7.8cm]{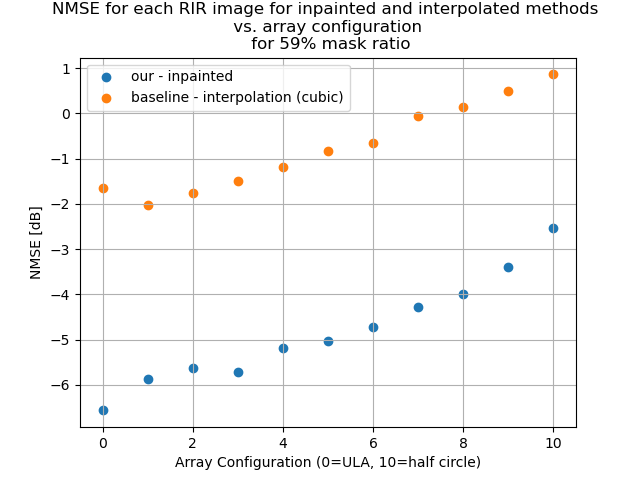}}}
 \caption{\ac{NMSE} for different array curvatures.}
 \label{fig:linear_to_circular_nmse_error}
\end{figure}

\begin{figure}[h!]
 \centerline{\framebox{
 \includegraphics[width=7.8cm]{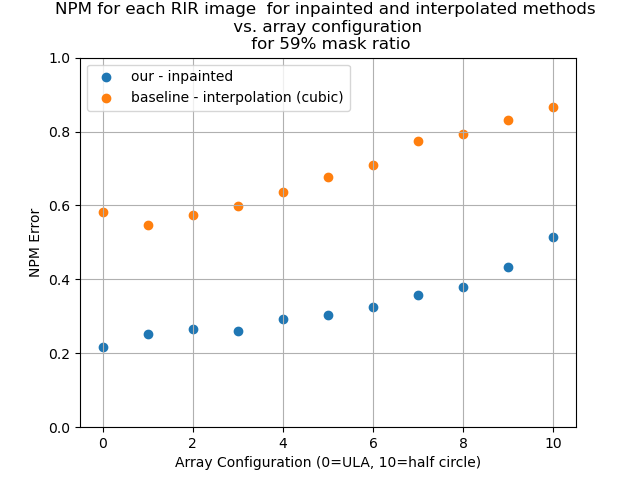}}}
 \caption{\ac{CD} for different array curvatures.}
 \label{fig:linear_to_circular_npm_error}
\end{figure}

The influence of array curvature on performance is further explored in Figs.~\ref{fig:linear_to_circular_nmse_error} and \ref{fig:linear_to_circular_npm_error}, which present the \ac{NMSE} and \ac{CD} measures, respectively. As the array curvature increases, performance degradation becomes evident, suggesting that the inpainting task is more manageable when the reflection patterns are straight rather than curved.
%
%

\begin{figure}[h!]
 \centerline{\framebox{
 \includegraphics[width=7.8cm]{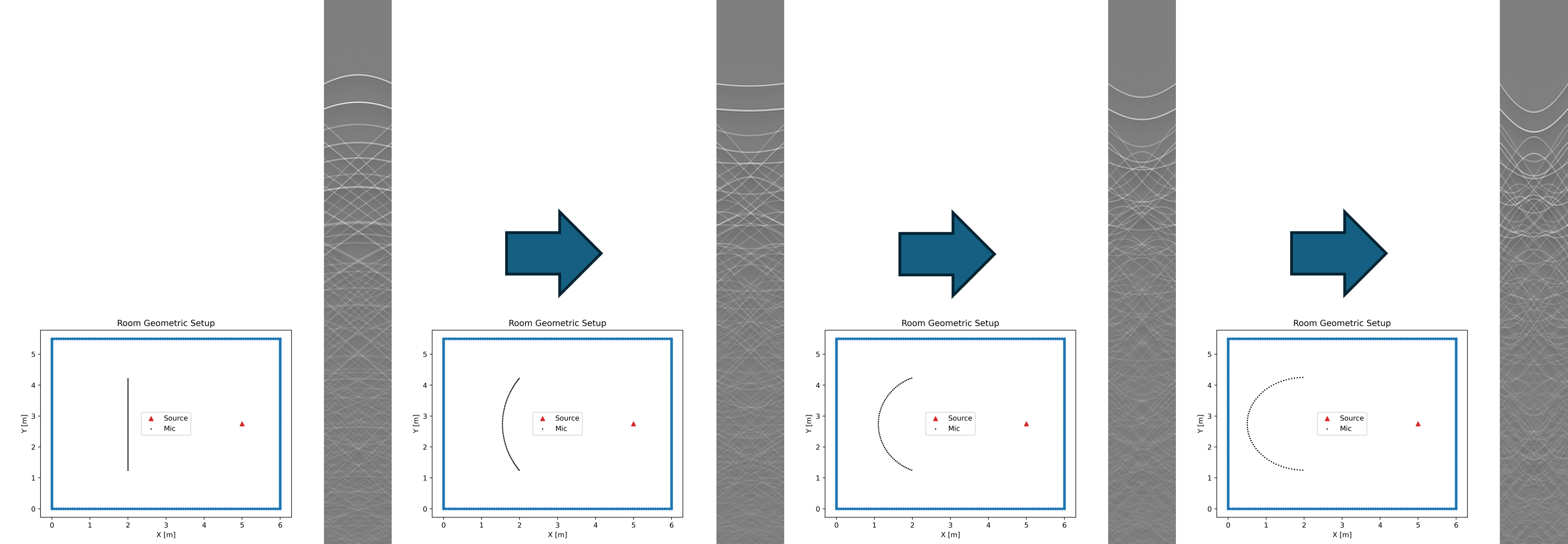}}}
 \caption{RIR images in grayscale for the linear array, circular array, and intermediate configurations.}
 \label{fig:linear_to_circular_rir_image}
\end{figure}

The \ac{RIR} images in grayscale for the linear array, the semi-circular array, and two intermediate configurations are presented in Fig.~\ref{fig:linear_to_circular_rir_image}. These images highlight differences in reflection patterns: the linear array exhibits straight-line reflections that are easier to reconstruct, whereas the semi-circular array produces curved reflections, which pose greater challenges during inpainting. These findings emphasize that while the model adapts well to semi-circular arrays, it achieves superior performance when the reflection paths are straight.

\begin{figure}[h!]
 \centerline{\framebox{
 \includegraphics[width=7.8cm]{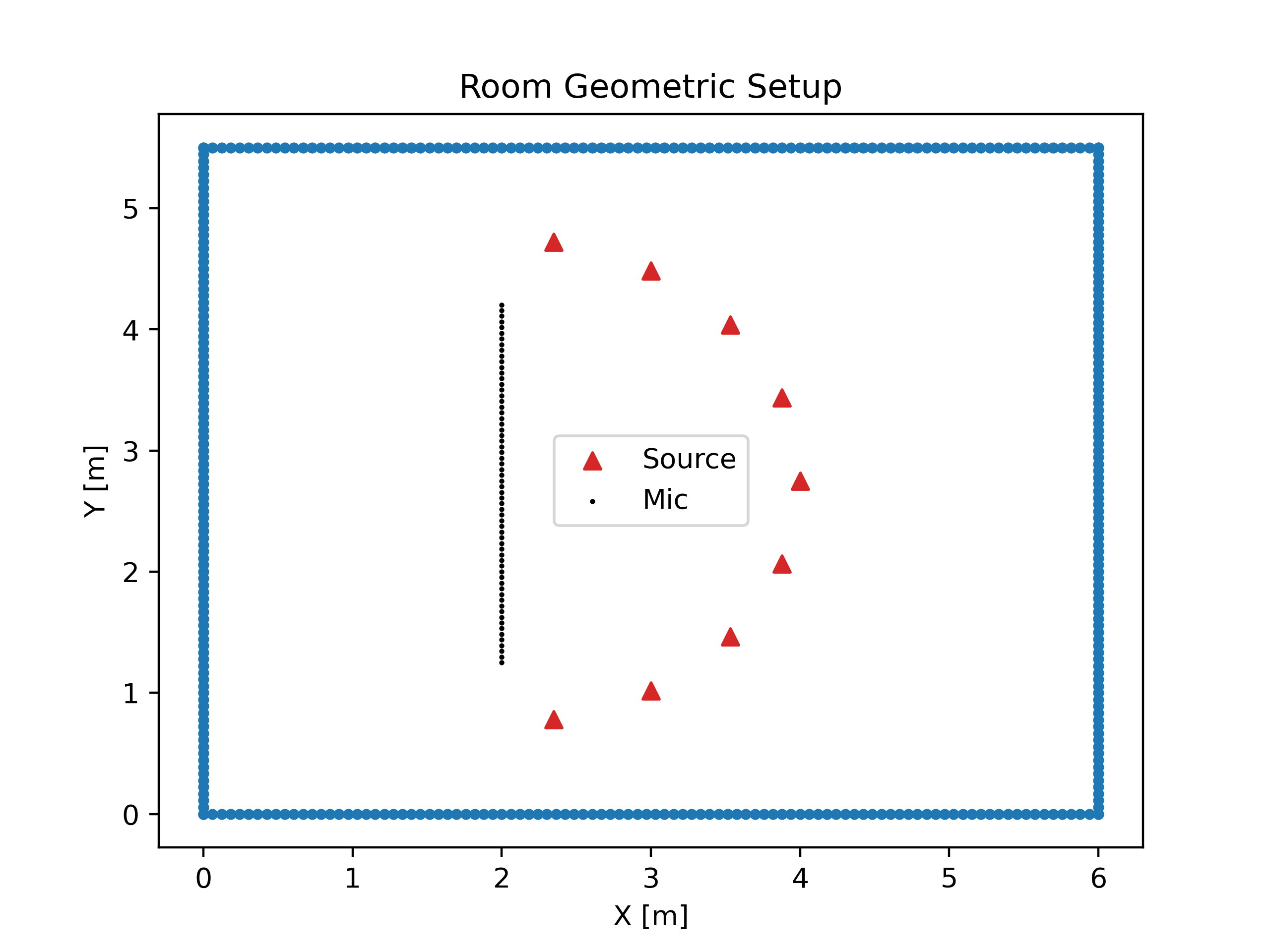}}}
 \caption{Geometric setup of the room with different source angles.}
 \label{fig:room_setup_different_source}
\end{figure}

\begin{figure}[h!]
 \centerline{\framebox{
 \includegraphics[width=7.8cm]{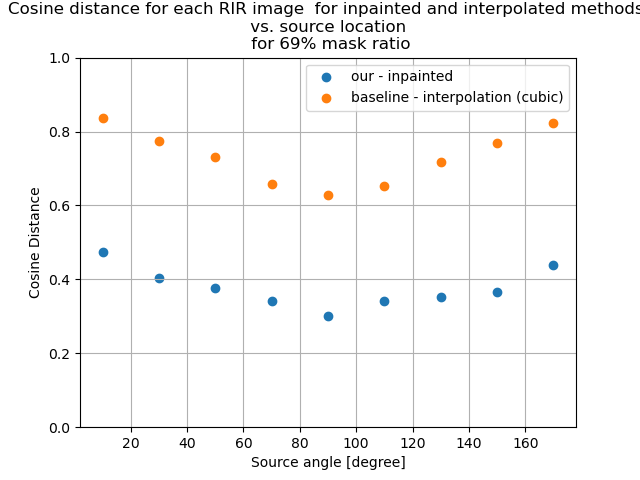}}}
 \caption{\ac{CD} for inpainted and baseline \ac{SCI} for different source location.}
 \label{fig:source_npm}
\end{figure}

Finally, in Fig.~\ref{fig:room_setup_different_source}, we show the room setup with 9 loudspeaker angles. Figure \ref{fig:source_npm} demonstrates that the best results are obtained for sources located at 90\textdegree{} (`broadside') relative to the array, while higher errors are observed for sources positioned at 10\textdegree{} or 170\textdegree{} (towards `endfire').

\begin{figure}[h!]
 \centerline{\framebox{
 \includegraphics[width=7.8cm]{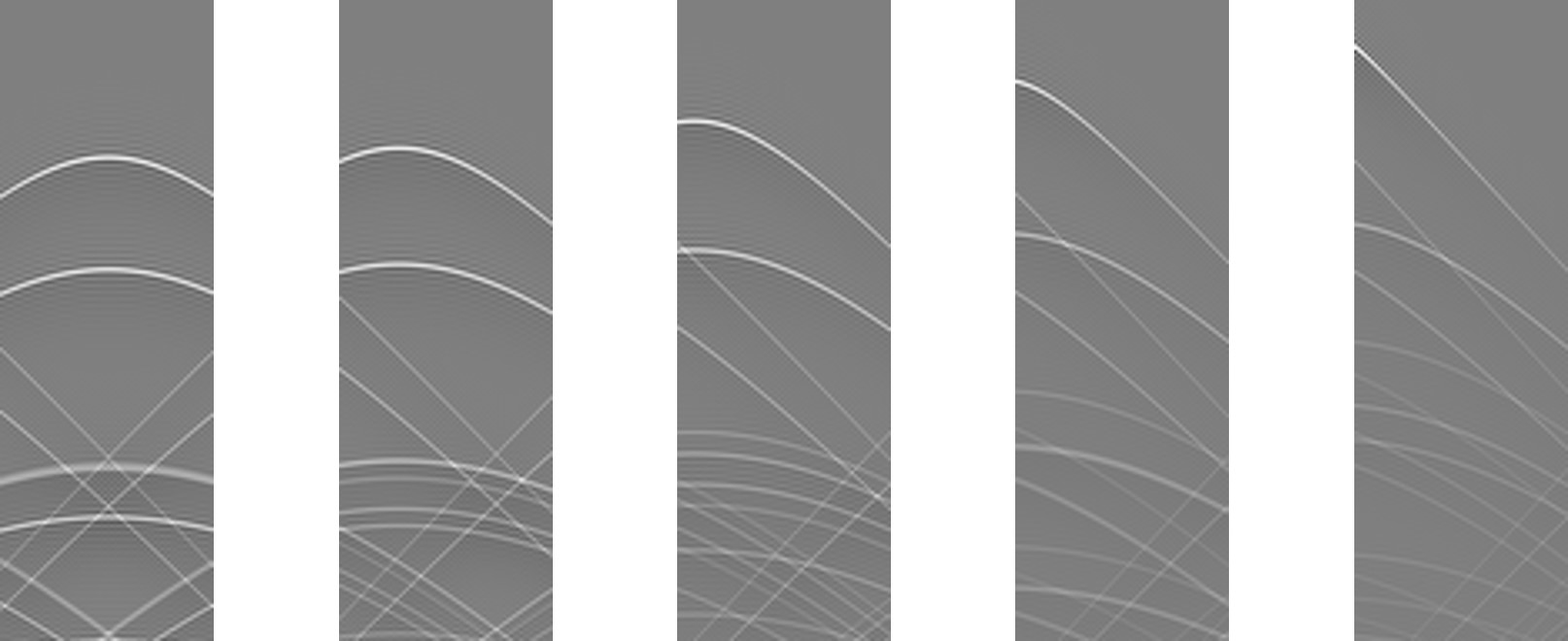}}}
 \caption{Comparison of \ac{RIR} images for different angles: from 90\textdegree{} - broadside (on the left) to 10\textdegree{} - endfire (on the right).}
 \label{fig:image_broadside_endfire}
\end{figure}

To further illustrate this, Fig.~\ref{fig:image_broadside_endfire} presents several grayscale images of \acp{RIR}, ranging from broadside to endfire configurations. When the source is positioned in front of the array, i.e., in the broadside configuration, the image exhibits greater symmetry, making it easier to inpaint and reconstruct the missing points. However, when the source is located at endfire angles, it becomes more challenging to complete the impulse response for the microphones on the opposite side, which are farther away. Additionally, the lines in the endfire configuration are sharper, while those from the broadside configuration are smoother.

\section{Discussion}

We addressed the challenge of acquiring \ac{RIR} measurements, which are essential for characterizing a room's acoustic properties but are resource-intensive to collect. We propose leveraging super-resolution techniques, traditionally used in imaging, to interpolate or predict \acp{RIR} at unmeasured locations within a room. This method utilizes existing \ac{RIR} data to generate high-resolution acoustic mappings without the need for exhaustive measurements, enabling applications in sound source localization, separation, and augmented reality.

Our simulation results show that the proposed method generalizes effectively beyond the trained configurations, allowing the generation of \acp{RIR} for different microphone arrays and even for rooms that were not part of the training set. Although tested with simulated \acp{RIR}, we believe that this research opens the door to generating additional data from limited real-world measurements.

\bibliography{bibliography}

\end{document}